\documentclass[letterpaper,12pt]{article}

\usepackage[T1]{fontenc}

\usepackage{dgjournal}
\usepackage{color}

\usepackage{mathptmx}

\usepackage{graphicx}

\usepackage[authoryear,comma,longnamesfirst,sectionbib]{natbib} 

\newtheorem{theorem}{Theorem}[section]

\begin{document}


~\\
{\bf \Large When is Menzerath-Altmann law mathematically trivial? A new approach}\\
~\\
Ramon Ferrer-i-Cancho$^1$, Antoni Hern\'andez-Fern\'andez$^{1,2}$, Jaume Baixeries$^1$, {\L}ukasz D\k{e}bowski$^3$ and J\'an Ma\v{c}utek$^{4}$\\
~\\
$^{1}$ Complexity \& Quantitative Linguistics Lab, LARCA Research Group, Departament de Llenguatges i Sistemes Inform\`atics, Universitat Polit\`ecnica de Catalunya, Campus Nord, Edifici Omega Jordi Girona Salgado 1-3, 08034 Barcelona (Catalonia), Spain\\
$^2$ Departament de Ling\"u\'{i}stica General, Universitat de Barcelona, Gran Via de les Corts Catalanes 585, 08007 Barcelona (Catalonia), Spain \\
$^3$ Institute of Computer Science, Polish Academy of Sciences, ul. Jana Kazimierza 5, 01-248 Warszawa, Poland\\
$^4$ Department of Applied Mathematics and Statistics, Comenius University, Mlynska dolina, 84248 
Bratislava, Slovakia\\
~\\
\begin{abstract}
Menzerath's law, the tendency of $Z$, the mean size of the parts, to decrease as $X$, the number of parts, increases is found in language, music and genomes. Recently, it has been argued that the presence of the law in genomes is an inevitable consequence of the fact that $Z = Y/X$, which would imply that $Z$ scales with $X$ as $Z \sim 1/X$. That scaling is a very particular case of Menzerath-Altmann law that has been rejected by means of a correlation test between $X$ and $Y$ in genomes, being $X$ the number of chromosomes of a species, $Y$ its genome size in bases and $Z$ the mean chromosome size. Here we review the statistical foundations of that test and consider three non-parametric tests based upon different correlation metrics and one parametric test to evaluate if $Z \sim 1/X$ in genomes. The most powerful test is a new non-parametric based upon the correlation ratio, which is able to reject $Z \sim 1/X$ in nine out of eleven taxonomic groups and detect a borderline group. Rather than a fact, $Z \sim 1/X$ is a baseline that real genomes do not meet. The view of Menzerath-Altmann law as inevitable is seriously flawed.
\end{abstract}
~
~\\
{\bf Corresponding author:} R. Ferrer-i-Cancho, rferrericancho@lsi.upc.edu

\section{Introduction}


Consider that $X$ and $Y$ are two discrete random variables and that $Z = Y/X$ with $X \neq 0$. For the particular case that $Z$ is a mean, Menzerath's law is the tendency of $Z$ to decrease as $X$ increases \citep{Menzerath1954a,Li2012}: $X$ stands for the number of parts of the construct (e.g., the number of clauses of a sentence), $Y$ stands for the size of the whole in parts (e.g., the length in words of the sentence) and $Z$ stands for the mean size of the construct (e.g., the mean length of the clauses in words). 
The scaling of $Z$ with $X$ that Menzerath's law describes qualitatively is typically modelled by means of 
Menzerath-Altmann law \citep{Altmann1980a,Cramer2005a}, i.e. 
\begin{equation}
Z = a X^be^{c X},
\label{Menzerath_Altmann_law_equation}
\end{equation}
where $a$, $b$ and $c$ are the parameters of the model.
Menzerath's law has been found in language \citep{Menzerath1954a} and genomes \citep{Ferrer2009e,Li2012} and also indirectly through a wide range of studies where Menzerath-Altmann law is fitted to human language e.g., \citep{Altmann1980a,Teupenhayn1984a}, and music \citep{Boroda1991a} and genomes \citep{Wilde1989a,Li2012} which yields parameters $a$, $b$ and $c$ that support a negative correlation between $Z$ and $X$ at least for sufficiently large $X$ (see \cite{Cramer2005a} for a review of parameter values). Note that in our definition of these laws, Menzerath's law is a light and model neutral law. In our view, the only constraint imposed by that law is that $Z$ and $X$ are negatively correlated (the dependency between $Z$ and $X$ might not be functional) while Menzerath-Altmann law is a strong assumption for two reasons: it assumes that the dependency between $Z$ and $X$ is functional and takes the form defined by Eq.\ \ref{Menzerath_Altmann_law_equation}. 

Recently, it has been argued that $Z = Y/X$ leads inevitably to a power-law of the form \citep{Sole2010a}
\begin{equation} 
Z = a X^{-1}, \label{Soles_law_equation}
\end{equation}
that is Menzerath-Altmann law with $b = -1$ and $c = 0$. If the argument was correct, Menzerath-Altmann law would be a trivial scaling law at least from a mathematical perspective. 
Being $X$ the number of chromosome of a species, $Y$ the genome size in bases of that species and $Z$ its mean chromosome size,
agreement with Eq.\ \ref{Soles_law_equation} has been claimed simply by fitting Menzerath-Altmann law with $c=0$ and obtaining $b\approx 1$ \citep{Sole2010a}. However, more than a decade of statistical research on presumable power laws in biology indicates that looks can be deceiving \citep{May2000a, Tanaka2005a,Stumpf2005a,Khanin2006a}. Sometimes, the divergence of the degree distribution of biological networks from a power-law is obvious upon visual inspection if a convenient representation of the data is employed 
\citep{Tanaka2005a}. In general, the hypothesis of a power-law for the degree distribution of biological networks has been rejected for different kinds of biological networks even when an exponentially truncated power-law similar to Eq.\ \ref{Menzerath_Altmann_law_equation} was considered \citep{Khanin2006a}. Modern model selection methods indicate that simple power-law models do not provide an adequate description of the degree distribution of protein interaction and metabolic networks \citep{Stumpf2005a,Stumpf2005b}. The same has happened to other hypotetical power-laws after careful inspection \citep{May2000a,Tjorve2003a}.
The dependency between the number of different species as a function of area is described more accurately by functions that are not power-laws \citep{May2000a}. Interestingly, the power-law model (as well as the exponential model) is seen as lacking biological depth for describing the species-area relationship with regard to other possible functions \citep{Tjorve2003a}. In a similar vein, Eq.\ \ref{Soles_law_equation} has been argued to imply assumptions that jeopardize chromosome well-formedness in the context of the relationship between chromosome number and mean chromosome length \citep{Baixeries2012a,Ferrer2012f}. This suggests that rather than indicators of complexity, pure power laws (or pure power-laws with a certain exponent, e.g., -1 in Eq.\ \ref{Soles_law_equation}) might play the role of baselines in certain circumstances.

The main goal of this article is presenting a powerful and statistically rigorous methodological framework to test if a real sample follows the particular case of power-law defined by Eq.\ \ref{Soles_law_equation}. Notice that the equation is a particular case of power-law for two reasons: the exponent of the power-law is $-1$ and, more importantly, the response variable $Z$ is $Z = Y/X$, being $X$ the predictor.

Whether Eq.\ \ref{Soles_law_equation} holds in genomes for the relationship between chromosome number and mean chromosome size,  has been debated \citep{Sole2010a,Ferrer2012f,Hernandez2011a,Baixeries2012a,Ferrer2012g,Baixeries2012b}. This is part of a long-running debate on the depth and importance of statistical laws of language in science (e.g., \cite{Miller1963,Li1992b,Suzuki2004a,McCowan2005a,Sole2010a,Ferrer2012f}). These laws are seen by many as inevitable \citep{Miller1968a,Sole2010a}, useless \citep{Suzuki2004a} or lacking mechanistic sophistication \citep{Li1992b,Stumpf2012a}. Here we aim to contribute to this general debate from the perspective of Menzerath-Altmann law with new theoretical insights and new experiments on genomes. In particular, we will provide some theoretical foundations for testing if Eq.\ \ref{Soles_law_equation} holds. It will be shown that rejecting Eq.\ \ref{Soles_law_equation} if $X$ and $Y$ are correlated \citep{Baixeries2012a,Hernandez2011a} is correct but conservative.
Furthermore, a new test that rejects Eq.\ \ref{Soles_law_equation} in all taxonomic groups considered so far except two (one being a borderline case) will be presented. The view of Menzerath-Altmann law as inevitable \citep{Sole2010a} is seriously flawed.

\section{Statistical foundations}

\subsection{The meaning of $Z = a/X$}

According to standard modelling, claiming that $Z$ scales with $X$ following Eq.\ \ref{Soles_law_equation} can be recast as \cite[pp. 1]{Ritz2008a},
\begin{equation}
E(Z | X = x) = a/x,
\end{equation}
for any $x$, being $E(Z | X = x)$ the conditional expectation of $Z$ given $x$ (a concrete value of $X$).
Testing if Eq.\ \ref{Soles_law_equation} holds reduces to testing if $Y$ is mean independent of $X$ \cite[pp. 67]{Poirier1995a}, namely $E(Y | X = x) = E(Y)$ for any $x$ \citep{Ferrer2012g}. Formally, this is supported by the following theorem \citep{Ferrer2012g}
\begin{theorem}
Consider a constant $a$ and two random natural variables, $X$ and $Y$, and a third random number $Z$, such that $X>0$ and $Z = Y/X$. Then,
$E(Z | X = x) = a/x$ if and only if 
$Y$ is mean independent of $X$, i.e. $E(Y | X = x) = E(Y)$ for any $x$. 
\label{mean_independence_theorem}
\end{theorem}
Therefore, one condition for a trivial Menzerath-Altmann law is that $Y$ is mean independent of $X$. Mean independence is well-known in econometrics \citep{Cameron2009a,Wooldridge2010a}.
Another more obvious mathematically trivial version of the law occurs when $Z$ is mean independent of $X$, i.e.
\begin{equation}
E(Z | X = x) = E(Z),
\end{equation}
which is equivalent to constant $E(Z| X = x)$ \citep{Ferrer2012g}. The analysis of the correlation between $Z$ and $X$ in genomes discarded this mean constancy of $Z$ for nine out of eleven taxonomic groups \citep{Ferrer2009e} and the result was confirmed with an updated dataset \citep{Baixeries2012a}. Therefore, Menzerath-Altmann law as a model of $E(Z | X = x)$ has two trivial versions: 
\begin{itemize}
\item
$b=c=0$: $Z$ is mean independent of $X$.
\item
$b=-1$ and $c=0$: $Y$ is mean independent of $X$. 
\end{itemize}
Interestingly, $b$ lies between $0$ and $-1$ when $c=0$ is assumed: e.g., $b=-0.27 \pm 0.11$ in language, being $Z$ is the mean clause length in sentences and $X$ is the number of sentences \citep{Teupenhayn1984a}, and $b=-0.44 \pm 0.09$ in music, being $Z$ is the mean F-motif length in tones and $X$ is the number of F-motifs \citep{Boroda1991a}. In both cases, we report $b = \mu \pm \sigma$, where $b$ is the exponent of a sample while $\mu$ and $\sigma$ are, the mean and the standard deviation of $b$ in an ensemble of samples, respectively.

\subsection{Three definitions of lack of association between $X$ and $Y$}

For the remainder of sections, it is important to bear in mind the definition of three statistical relations between $X$ and $Y$ \cite[pp. 67-68]{Poirier1995a}:
\begin{itemize}
\item
{\em $X$ and $Y$ are independent:} $p(Y = y | X = x) = p(Y= y)$ for any $x$ and $y$.
\item
{\em $Y$ is mean independent of $X$: } $E(Y | X = x) = E(Y)$ for any $x$. 
\item
{\em $X$ and $Y$ are Pearson uncorrelated: } $COV(X,Y) = 0$ where $COV(X,Y) = E(XY)- E(X)E(Y)$ is the covariance between $X$ and $Y$. Notice that Pearson uncorrelation, i.e. $\rho(X,Y) = 0$, being $\rho(X,Y)$ the Pearson correlation coefficient, is equivalent to zero covariance as \citep{DeGroot2012a}
\begin{equation} 
\rho(X,Y) = \frac{COV(X,Y)}{\sigma(X) \sigma(Y)},
\end{equation}
with $\sigma(X)$ and $\sigma(Y)$ as the standard deviation of $X$ and $Y$, respectively. 
\end{itemize}
As $X$ and $Y$ are Pearson uncorrelated if and only if $\rho(X,Y)=0$ (or $COV(X,Y) = 0$), $Y$ is mean independent of $X$ if and only if $\eta(Y, X)=0$, where $\eta(Y, X)$ is a less-known association metric: the correlation ratio \citep{Crathorne1922a,Kruskal1958a}. $\eta(Y, X)$ derives from the variance of $E(Y | X = x)$, which is by definition, 
\begin{equation}
Var(E(Y | X = x)) = E([E(Y|X =x) - E(E(Y| X = x))]^2).
\end{equation}
By the law of total probability for expectations \cite[pp. 258]{DeGroot2012a}, $E(E(Y| X = x)) = E(Y)$ and thus
\begin{equation}
Var(E(Y | X = x)) = E([E(Y|X =x) - E(Y)]^2).
\end{equation}
From this variance, the correlation ratio of $Y$ on $X$ is defined as \citep{Kruskal1958a}
\begin{equation}
\eta(Y, X) = \left[ \frac{Var(E(Y | X = x))}{Var(Y)} \right]^{1/2} = \frac{\sigma(E(Y | X = x))}{\sigma(Y)}, 
\end{equation}
where $\sigma(...)$ indicates the standard deviation. Notice that $0 \leq \eta(Y, X) \leq 1$ whereas $-1 \leq \rho(X, Y) \leq 1$ \cite[pp. 816-817]{Kruskal1958a}. As $\rho(X,Y)$ is a normalized $COV(X,Y)$, $\eta(Y,X)$ is a normalized $Var(E(Y | X = x))$.
Interestingly, the correlation ratio satisfies the following properties \cite[pp. 816-817]{Kruskal1958a}:
\begin{itemize}
\item
$\eta(Y, X) = 1$ if and only if $Y$ is a perfect function of $X$.
\item
$|\rho(X,Y)| \leq \eta(Y, X)$ with equality if and only if $Y$ is a linear function of $X$.
\item
$0 \leq \eta(Y, X) \leq 1$ (whereas $-1 \leq \rho(X, Y) \leq 1$). As $\rho(X,Y)$ is a normalized $COV(X,Y)$, $\eta(Y,X)$ is a normalized $Var(E(Y | X = x))$.
\end{itemize}

It is well-known that \citep{Kolmogorov1956a,Poirier1995a}: \\
\begin{eqnarray}
\begin{array}{c}
X \mbox{~and~} Y \mbox{~are independent} \\ 
\Downarrow \\ 
Y \mbox{~is mean independent of~} X \mbox{~~}(\eta(Y,X) = 0)\\ 
\Downarrow \\
X \mbox{~and~} Y \mbox{~are Pearson uncorrelated~} \mbox{~~}(\rho(X,Y) = 0)
\end{array}
\nonumber
\end{eqnarray}
Proofs of the top to bottom implications have been provided by, e.g., \cite[pp. 60]{Kolmogorov1956a} or \cite[pp. 67]{Poirier1995a}.
Mean independence implies Pearson uncorrelation but the converse (Pearson uncorrelation implies mean independence) is not necessarily true. To see it consider that 
\begin{equation}
p(X = x, Y = y) = \left\{ 
   \begin{array}{cc} 
      1/2 & \mbox{if~} x = 0 \mbox{~and~} y = -1 \\
      1/4 & \mbox{if~} x = -1 \mbox{~and~} y = 1 \\
      1/4 & \mbox{if~} x = 1 \mbox{~and~} y = 1
   \end{array}
\right.
\end{equation}
Thus $E(X)=E(Y)=0$ and $E(XY)=(1/4)(-1) + (1/2)0 + (1/4)1 = 0$. Therefore $COV(X,Y)=0$ but $Y$ is not mean independent of $X$ because $E(Y|X=-1)=1 \neq E(Y|X=0)=-1$. Similarly, independence implies mean independence but the converse (mean independence implies independence) is not necessarily true (see \cite{Ferrer2012g} for a counterexample).

In next section it will be shown that the correlation ratio is indeed more powerful than a correlation coefficient for testing whether Eq.\ \ref{Soles_law_equation} holds.

\section{How to test that $Z=a/X$}

\subsection{Non-parametric tests}

Here we consider three sample correlation statistics: the Pearson correlation $\rho(X,Y)$, the Spearman correlation $\rho_S(X,Y)$ and the correlation ratio $\eta(X,Y)$, and evaluate if they are significantly different from zero using a permutation test (a particular case of randomization test \cite[pp. 803-819]{Sokal1995}).
In a sample of size $N$, $X$ and $Y$ can be seen as vectors, i.e. $X = \{x_1,...,x_i,...,x_N\}$ and $Y = \{y_1,...,y_i,...,y_N\}$, being $(x_i,y_i)$ the information associated to the $i$-th element of the sample. 
The $p$-value of $\rho(X,Y)$ is the probability that $X'$, a random permutation of $X$, yields $|\rho(X',Y)| \geq |\rho(X,Y)|$. The test is two-sided as both positive and negative values of the sample $\rho(X,Y)$ that are significantly different from 0 are indicative of non-zero correlation 
(both $\rho(X,Y)=1$ and $\rho(X,Y)=-1$ are indicative of maximum correlation). 
The $p$-value of $\rho_S(X,Y)$ is calculated as that of $\rho(X,Y)$.
The $p$-value of $\eta(Y, X)$ is the probability that $X'$ (a random permutation of $X$) yields $\eta(Y, X') \geq \eta(Y,X)$. The test is one-sided as 
only positive values of the sample $\eta(X,Y)$ that are significantly large are indicative of non-zero correlation
($\eta(X, Y)$ is indicative of maximum correlation only when $\eta(X, Y) = 1$).
The $p$-values were estimated with a Monte Carlo method generating $R = 10^7$ uniformly random permutations. In the present article, we adopt a significance level of 0.05.

\begin{table}
\begin{center}
\caption{\label{statistical_properties_table} A summary of the statistical properties of each taxonomic group. $N$ is the number of species. The mean ($\mu$) and sample standard deviation ($\sigma$) are provided for three variables: chromosome number ($X$), genome size ($Y$) in Million base pairs (Mb) and mean chromosome size ($Z$) in Mb per chromosome. The genome size in Mb is measured using the DNA $C$-value ($1C$). Accordingly, the number of chromosomes is measured using $1n$. Mean and standard deviation were rounded to leave just one decimal digit.}

\begin{tabular}{@{\extracolsep{4pt}}lrcccccc}
\hline\noalign{\smallskip}
                & & \multicolumn{2}{c}{$X$} & \multicolumn{2}{c}{$Y$} & \multicolumn{2}{c}{$Z$} \\
                \cline{3-4} \cline{5-6} \cline{7-8}
Taxonomic & $N$ & $\mu$ & $\sigma$ & $\mu$ & $\sigma$ & $\mu$ & $\sigma$ \\
group     &     &       &          &       &          &       &          \\  
\noalign{\smallskip}\hline\noalign{\smallskip}
Fungi & 56 & 9.4 & 4.2 & 33.3 & 23.7 & 3.8 & 1.9 \\
Angiosperms & 4706 & 14.2 & 9.6 & 5474.3 & 8902.5 & 515 & 847 \\
Gymnosperms & 170 & 12 & 2.3 & 17695.4 & 7436.9 & 1489.6 & 613.3 \\
Insects & 269 & 10.6 & 5 & 1240 & 1908.7 & 146 & 204.3 \\
Reptiles & 170 & 19.1 & 3.8 & 2364.2 & 742.6 & 126 & 39.9 \\
Birds & 99 & 38.8 & 31.8 & 1400.7 & 175 & 159.9 & 243.1 \\
Mammals & 371 & 23.3 & 8.8 & 3413.2 & 822.1 & 182.1 & 134.2 \\
Cartilaginous & 52 & 36.7 & 8.2 & 5787.9 & 3100.4 & 166.5 & 95.1 \\
fishes        &    &      &     &        &        &       &      \\  
Jawless fishes & 13 & 44 & 33.7 & 2442.7 & 1090.4 & 140.7 & 174.7 \\
Ray-finned & 647 & 27.6 & 10.7 & 1360.7 & 716.9 & 50.7 & 27.7 \\
fishes        &    &      &     &        &        &       &      \\  
Amphibians & 315& 14.2 & 5.5 & 18029 & 19952.3 & 1285.3 & 1373.4 \\
\noalign{\smallskip}\hline
\end{tabular}
\end{center}
\end{table}

For the analyses of this section, we used the same dataset of recent research \citep{Hernandez2011a,Baixeries2012a,Baixeries2012b,Ferrer2012f}. A summary of the statistical properties of the dataset is provided in Table \ref{statistical_properties_table}. See \cite{Ferrer2009e,Hernandez2011a} for further details about the dataset.
As the fact that $Y$ is mean independent of $X$ implies Pearson uncorrelation, i.e. $COV(X,Y) = 0$ (or $\rho(X,Y) = 0$), Eq.\ \ref{Soles_law_equation} can be tested by means of the following procedure: if $COV(X,Y)$ (or $\rho(X,Y) = 0$) is significantly different from $0$ then reject Eq.\ \ref{Soles_law_equation}, otherwise accept it.
That procedure can be used to reject Eq.\ \ref{Soles_law_equation} in genomes, being $Z$ the mean chromosome length in bases of a species and $X$ being the number of chromosomes of that species \citep{Hernandez2011a}. 

Table \ref{correlation_test_table} summarizes the results of the analysis of the Pearson correlation between $X$ (chromosome number) and $Y$ (genome size in bases) in genomes. A significant correlation is found in six out of eleven taxonomic groups. The test is conservative as it rejects Eq.\ \ref{Soles_law_equation} indirectly by means of a necessary condition for this equation to hold: $COV(X,Y) = 0$. Therefore, the five groups where the Eq.\ \ref{Soles_law_equation} could not be rejected might be false negatives. 
Pearson correlation is a measure of linearity between variables and has difficulties for capturing non-linear dependencies. A possible improvement is using a more powerful correlation metric such as $\rho_S(X,Y)$, the Spearman rank correlation coefficient, which is a measure of monotonic (linear or non-linear) dependency \citep{Zou2003a}. The Spearman rank correlation is a Pearson correlation on a rank transformation of the original values \citep{Conover1999a}.
The Spearman rank correlation test revealed that the majority of taxonomic groups (nine out of eleven) exhibit a significant correlation between $X$ and $Y$ that is incompatible with Eq.\ \ref{Soles_law_equation} (Table \ref{correlation_test_table}). The exceptions are birds and cartilaginous fishes.
These findings confirm qualitatively the previous results with similar methods \citep{Hernandez2011a}. Interestingly, there is a more powerful way of testing Eq.\ \ref{Soles_law_equation}: testing directly if $Y$ is mean independent of $X$ from its definition, i.e. $E(Y|X= x) = E(Y)$ for any $x$, which is equivalent to $\eta(X,Y) = 0$ \citep{Kruskal1958a}. 
$\eta(Y, X)$ was not significantly large in two taxonomic groups: cartilaginous fishes and jawless fishes but the case of jawless fishes is border line: the estimated p-value is $0.054$. The fact that this is the group with the smallest sample ($N=13$), suggests that this might be a case of type II error.

\begin{table*}
\begin{center}
\caption{\label{correlation_test_table} 
Analysis of the correlation between $Y$ (genome size in bases) and $X$ (chromosome number). $D(X)$ is the proportion of different values of $X$. $\rho(X,Y)$ is the sample Pearson correlation coefficient, $\rho_S(X,Y)$ is the sample Spearman correlation coefficient and 
$\eta(Y,X)$ is the sample correlation ratio of $Y$ on $X$. The sample size corresponds to $N$ in Table \ref{statistical_properties_table}.
The $p$-values were estimated using a Monte Carlo permutation test.
The correlation statistics and the corresponding $p$-values were rounded to leave only two significant digits.
}
\begin{tabular}{llllllll}
\hline\noalign{\smallskip}
Taxonomic & $D(X)$ & $\rho(X,Y)$ & $p$-value & $\rho_S(X,Y)$ & $p$-value & $\eta(Y, X)$ & $p$-value \\
group \\
\noalign{\smallskip}\hline\noalign{\smallskip}
Fungi & $0.25$ & $0.41$ & $0.002$ & $0.28$ & $0.037$ & $0.74$ & $0.026$ \\
Angiosperms & $0.02$ & $-0.0024$ & $0.87$ & $-0.039$ & $0.008$ & $0.27$ & $0.0013$ \\
Gymnosperms & $0.08$ & $0.1$ & $0.18$ & $0.32$ & $2.9 \cdot 10^{-5}$ & $0.74$ & $<10^{-7}$ \\
Insects & $0.08$ & $0.09$ & $0.14$ & $0.22$ & $0.0003$ & $0.46$ & $0.017$\\
Reptiles & $0.11$ &$0.31$ & $3.6 \cdot 10^{-5}$ & $0.24$ & $0.0014$ & $0.48$ & $0.0031$ \\
Birds & $0.36$ &$-0.029$ & $0.78$ & $-0.033$ & $0.74$ & $0.78$ & $0.0012$ \\
Mammals & $0.11$ &$0.3$ & $<10^{-7}$ & $0.3$ & $<10^{-7}$ & $0.68$ & $<10^{-7}$ \\
Cartilaginous & $0.44$ &$0.014$ & $0.92$ & $-0.13$ & $0.36$ & $0.76$ & $0.11$ \\
fishes \\
Jawless fishes & $0.77$ &$-0.76$ & $0.0033$ & $-0.74$ & $0.0049$ & $0.98$ & $0.054$ \\
Ray-finned & $0.07$ & $0.47$ & $<10^{-7}$ & $0.49$ & $<10^{-7}$ & $0.69$ & $<10^{-7}$ \\
fishes \\
Amphibians & $0.07$ & $0.13$ & $0.02$ & $0.45$ & $<10^{-7}$ &$0.58$ & $<10^{-7}$ \\
\noalign{\smallskip}\hline
\end{tabular}
\end{center}
\end{table*}

\begin{figure}[!h]
\centering
\includegraphics[scale = 0.6]{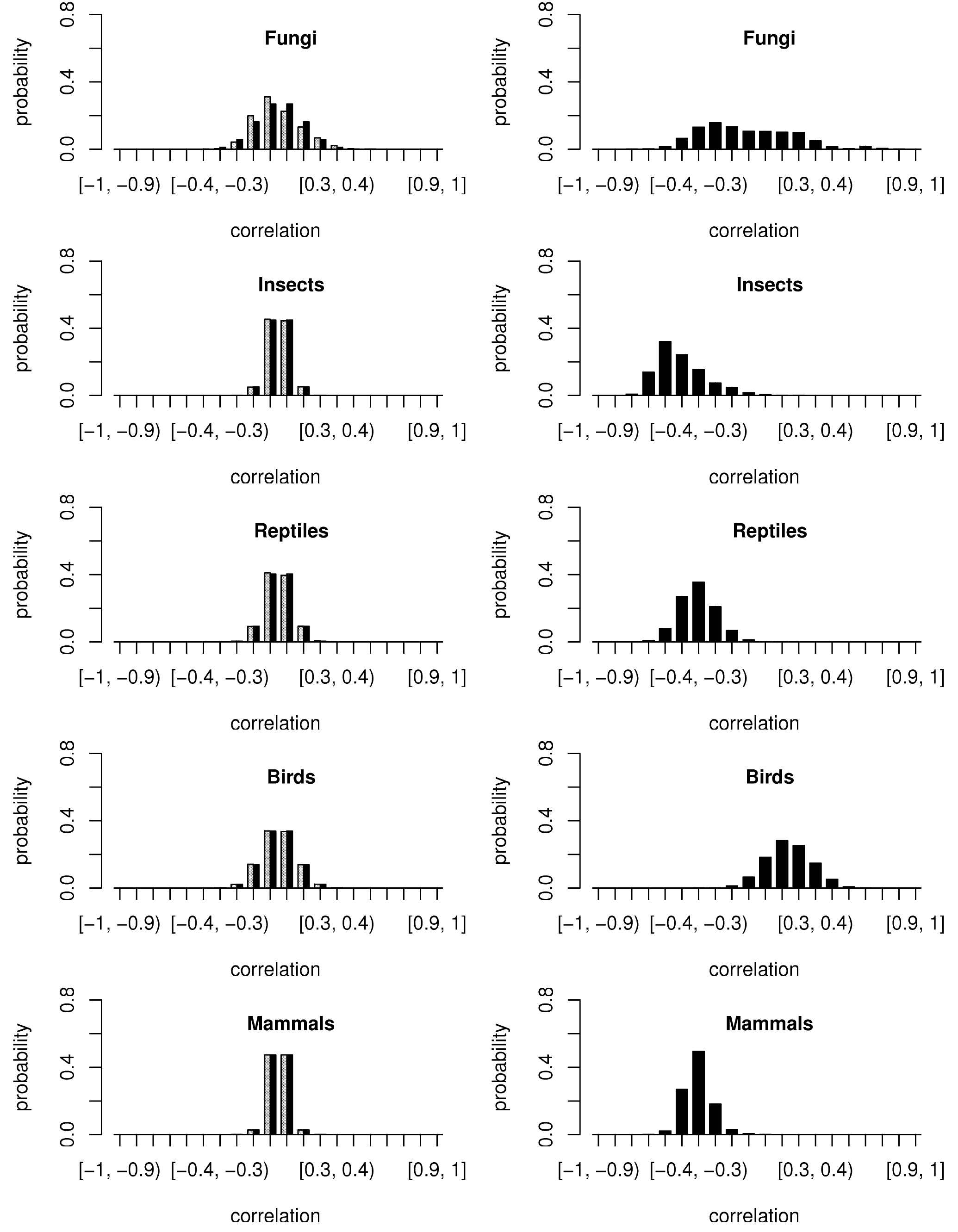}
\caption{\label{histograms1_figure} Normalized histograms of the correlations statistics. The range of variation of every statistics is split into 20 bins. Left column: histograms of the Pearson (black) and Spearman (gray) rank correlation with a bin size of 0.1. Right column: histogram of the correlation ratio with a bin size of 0.05. $10^7$ randomizations were used. }
\end{figure}

\begin{figure}[!h]
\centering
\includegraphics[scale = 0.6]{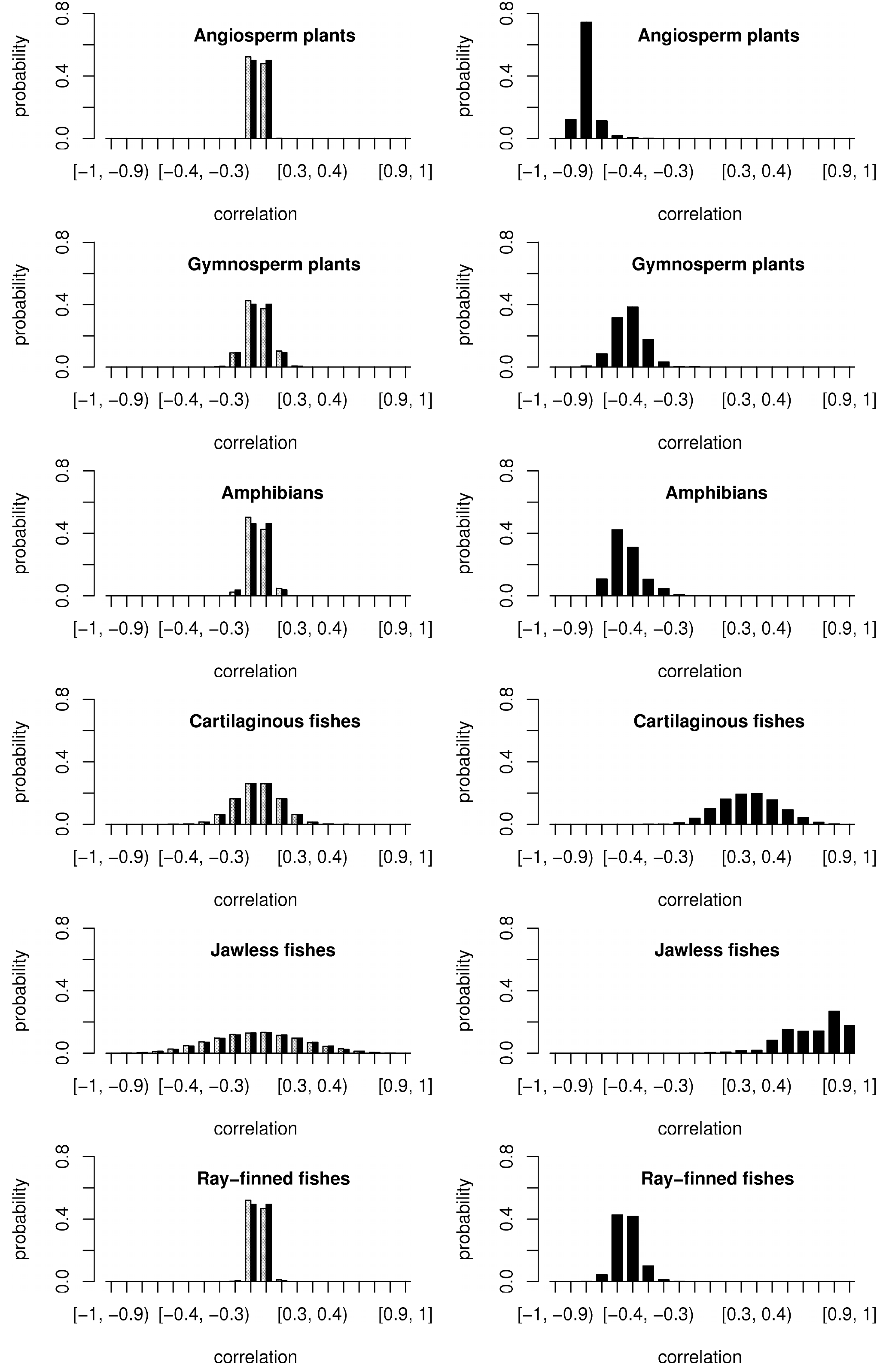}
\caption{\label{histograms2_figure} The same as in Table \ref{histograms1_figure} for other taxonomic groups. }
\end{figure}

Fig. \ref{histograms1_figure} and Fig. \ref{histograms2_figure} show histograms of the value of three correlation statistics in randomizations of the original data. As expected, the distribution for the Pearson and Spearman correlation statistics is centered at zero and symmetric. Fig. \ref{histograms2_figure} reveals crucial information to understand the taxonomic groups where the correlation ratio fails to reject the null hypothesis: the density of probability is concentrated on high values of the statistic. The case of jawless fishes is particularly intriguing: why the null hypothesis cannot be rejected when the correlation ratio is almost maximum? Table \ref{correlation_test_table} shows that $\eta(Y,X) = 0.98$ for jawless fishes while the maximum is $\eta(Y,X) = 1$). We believe that there is not a biological but a statistical reason for that: the test based on $\eta(Y,X)$ loses power (capacity to reject the null hypothesis) when the diversity of values of $X$ is high. If $\phi(X)$ is defined as the set of values of $x_i$ in the sample of a taxonomic group, i.e.
\begin{equation}
\phi(X) = \cup_{i=1}^N \{x_i\}
\end{equation}
then one can define a simple diversity index $D(X)$ as the proportion of different values of $X$, i.e.  
\begin{equation}
D(X) = \frac{|\phi(X)|}{N}.
\label{diversity_ratio_equation}
\end{equation}
Interestingly, the two groups with the largest $D(X)$ coincide with the two groups where the correlation ratio test fails to reject the null hypothesis (Table \ref{correlation_test_table}). It will shown that $\eta(Y,X)$ is maximum, i.e. $\eta(Y,X) = 1$, when the diversity of $X$ is maximum, i.e. $D(X)= 1$, regardless of the values of $Y$. Let us define $\phi(Y|w)$ as the set of values of $y_i$ such that $x_i = w$, i.e.
\begin{equation}
\phi(Y|w) = \cup_{x_i = w, 1 \leq i \leq N} \{y_i\}
\end{equation}
and use $Var(...)$ to refer (in this context) to the sample variance and $\mu(Y|w)$ as the mean of $Y$ conditioned on $w$ ($\mu(Y|w)$ is the mean value of $y_i$ for $(x_i,y_i)$ pairs such that $x_i = w$).
The sample correlation ratio can be defined as function of means and sample variances as 
\begin{equation}
\eta(Y,X) = \left( \frac{Var(\mu(Y|x))}{Var(Y)} \right)^{1/2}
\end{equation}
If $D(X)=1$, i.e. $|\phi(X)| = N$ (Eq. \ref{diversity_ratio_equation}), then $|\phi(Y|x)|=1$ for every $x$ in $\phi(X)$. This has a very important consequence, namely that $\mu(Y|x_i) = y_i$, which yields $Var(\mu(Y|x)) = Var(Y)$ and therefore $\eta(Y,X)=1$.

The fact that $\eta(Y,X)=1$ when there are not repeated values of $X$ means that $\eta(Y,X)$ will keep being maximum for any randomization of the data and thus the p-value of $\eta(Y,X)$ is 1. Put differently, the data might be showing a perfect correlation between $X$ and $Y$ but the absence of repeated values of $X$ can lead to a type II error (false negative) when $D(X)=1$. We believe that this can also happen when $D(X)<1$ but $D(X)$ is sufficiently large. Not surprisingly, the only two groups where the correlation ratio test fails to reject the null hypothesis at a significance level of 0.05, i.e. cartilaginous fishes and jawless fishes, are the two groups with the largest $D(X)$ ($D(X)= 0.44$ and $D(X)= 0.77$, respectively; Table \ref{correlation_test_table}).
$D(X)$ allows one to solve the puzzle of why the density of probability of the correlation ratio is concentrated on high values of $\eta(Y,X)$, specially for cartilaginous fishes and jawless fishes (Fig. \ref{histograms2_figure}): their high value of $D(X)$ is biasing $\eta(Y,X)$ towards high values in randomization of the original data, no matter how the values of $X$ and $Y$ combine under the null hypothesis.

\subsection{A parametric test}

The hypothesis of $Z=a/X$ has been accepted with the only support that the fit of $Z = aX^b$ yields $b \approx -1$ \citep{Sole2010a}. This procedure is very prone to type II error (accepting a false null hypothesis) as it needs that 
$Z = aX^b$ holds first \citep{Baixeries2012b,Ferrer2012f}. Our analysis shows that for $Z=a/X$ to hold, it is not only necessary that $b \approx -1$ is retrieved but also $a \approx \mu(Y)$, where $\mu(Y)$ is the mean of $Y$, an estimate of $E(Y)$ (recall Theorem \ref{mean_independence_theorem}; see also \citep{Ferrer2012g}). Even if $a \approx \mu(Y)$ and $b \approx -1$, type II errors are not excluded and minimizing them needs evidence that $Z = aX^b$ is well-supported by data.

\section{Discussion}

We have argued that claiming that $Z$ scales with $X$ following a very particular form of Menzerath-Altmann law, i.e. Eq.\ \ref{Soles_law_equation}, is indeed equivalent to claiming that $E(Z | X) = E(Y)/X$, which is indeed equivalent to claiming that $Y$ is mean independent of $X$. We have also presented a new correlation ratio test revealing that Eq.\ \ref{Soles_law_equation} could only hold in cartilaginous fishes and maybe also in jawless fishes. The two major conclusion of our article are that 
\begin{itemize}
\item
The trivial scaling defined by Eq.\ \ref{Soles_law_equation} is the exception, not the rule.
\item
The correlation ratio is at least one of the correlation statistics to consider for testing the null hypothesis that $Z = a/X$.
\end{itemize}

While the Pearson correlation test rejects Eq.\ \ref{Soles_law_equation} in 6 taxonomic groups, the Spearman correlation and the correlation ratio tests do it in 9 taxonomic groups (Table \ref{correlation_test_table}).   
The two most powerful tests might be failing because of type II errors. This possibility is apparently easier to defend for the correlation ratio from the perspective of insufficient sampling: the correlation ratio test fails to reject the null hypothesis exactly for the two groups having the smallest sample size (jawless fishes and cartilaginous fishes). In contrast, the Spearman correlation test fails with the second and fourth taxonomic groups with the smallest sample size, i.e. jawless fishes and birds (recall Table \ref{statistical_properties_table}). A deeper explanation of the failure of the correlation ratio is offered by $D(X)$. 

We have shown that the power of the correlation test depends on the diversity of values of $X$, losing power completely if that diversity is maximum.
The two groups where that statistic fails to reject the null hypothesis are not only the two groups with maximum $D(X)$ but also the two groups with the smallest $N$ (Table \ref{statistical_properties_table} and \ref{correlation_test_table}). The latter suggest that the true reason for the failure is that the  samples of these two groups are not large enough for significance to emerge. Indeed, $D(X)$ is a function of $N$. 
However, if a small sample size was the only reason for the failure to reject the null hypothesis, one would also expect that fungi (with $N=56$) were also problematic, but their p-values are about half the significance level of 0.05. While the value of $N$ cartilaginous fishes and fungi are very similar (the value of $N$ of cartilaginous fishes is about $92\%$ of the value of $N$ of fungi; Table \ref{statistical_properties_table}), the differences between both groups in terms of $D(X)$ are much larger (the value of $D(X)$ of cartilaginous fishes is about $176\%$ the value of $D(X)$ of fungi; Table \ref{correlation_test_table}).   
Therefore, a high $D(X)$ is more likely to be the primary reason for the failures of the correlation ratio tests (notice that $N$ could be very large but the sample $\eta(Y,X)$ may not be significantly large because $D(X)=1$). 

Whether the correlation ratio is a more powerful statistic than the Spearman rank correlation must be investigated further.  
If the significant level is raised to 0.01, then it turns out that the Spearman rank correlation allows one to reject the null hypothesis in 8 taxonomic groups while the correlation ratio does it in 7 groups (Table \ref{correlation_test_table}). However, the difference is small and both statistics may need to be used jointly.

The fact that $Z = a/X$ does not hold in many taxonomic groups must be interpreted with care. The taxonomic groups are not totally independent and thus the number of groups in which a statistic rejects the null hypothesis might be uninformative: it may simply reflect the existence of a common ancestor in their evolutionary history (the same argument could be applied to the groups where the null hypothesis could not be rejected: their number might be uninformative). However, the hypothesis of a common ancestor is incomplete at present because 
\begin{itemize}
\item
A demonstration of the existence of such ancestor  is necessary. That demonstration would require a careful analysis of the transmission of traits throughout phylogeny (e.g., \cite{Dunn2011a}) 
and this is beyond the scope of this article.
\item
An explanation of why the common ancestor group adopted that property would still be required. 
\end{itemize}
However, the many rejections of the null hypothesis might be due to the convergence of distinct taxonomic groups to the same statistical pattern by some general constraints, maybe physical or biological principles. This alternative to the common ancestor is supported by our suspect that the reason why the correlation ratio has failed in two groups are not biological but rather statistical (type II errors).
The possibility that one of those general constraints is the viability and well-formedness of chromosomes is reviewed next.

The random breakage model where $X$ and $Y$ are independent and uniformly distributed \citep{Sole2010a} fails to fit the majority of taxonomic groups because independence is a particular case of mean independence \citep{Ferrer2012g} and mean independence fails in at least nine out of eleven taxonomic groups (Table \ref{correlation_test_table}). Interestingly, it has been argued that independence between $X$ and $Y$ is problematic as it can lead to organisms with empty chromosomes or empty chromosome parts \citep{Baixeries2012a}. Notice that $Y$ does not need to be the size of genome in bases. It could be the size in units between the base and the chromosome. 

The problem of empty components also concerns mean independence. The condition for not expecting empty chromosomes for a given $x$ (a concrete value of $X$) is 
\begin{eqnarray}
E(Z | X = x) & \geq & 1 \nonumber \\ 
\frac{1}{x} E(Y | X = x) & \geq & 1. \label{no_empty_chromosomes_equation}
\end{eqnarray}
For that $x$, the condition in Eq.\ \ref{no_empty_chromosomes_equation} becomes $E(Y) \geq x$ when $Y$ is mean independent of $X$ as $E(Y| X = x) = E(Y)$ in that case. Thus, under mean independence, empty chromosomes are expected in an organism of $x$ chromosomes if $E(Y) < x$. Notice that expecting that Eq.\ \ref{no_empty_chromosomes_equation} holds on average for any $x$, leads to
\begin{eqnarray}
E(E(Y | X = x)) & \geq & E(X) \nonumber \\
E(Y) & \geq & E(X) \label{no_empty_chromosomes_on_average_equation}
\end{eqnarray}
thanks to the law of total probability for expectations \cite[pp. 258]{DeGroot2012a}.
The restrictions defined by Eqs.\ \ref{no_empty_chromosomes_equation} and \ref{no_empty_chromosomes_on_average_equation} are perhaps very simple but \cite{Baixeries2012a} considered more elaborated constraints for the viability of an organism based upon the parts making an ideal chromosome: a centromere, two telomeres and a couple of intermediate regions. Those viability constraints lead to a deviation from Menzerath-Altmann law with $b=-1$ and $c=0$ (see Fig. 3 of \cite{Baixeries2012a}), which Theorem \ref{mean_independence_theorem} allows one to interpret unequivocally as a departure from mean independence. Therefore, the viability and well-formedness of chromosomes is not compatible with mean independence either. This suggests that a common ancestor might not be the only reason for the recurrent failure of $Z = a/X$ in taxonomic groups. Indeed, the existence of general and lineage independent constraints is a more parsimonious hypothesis that has to be investigated further.

The negative Spearman rank correlation between $Z$ and $X$, known as Menzerath's law, defies a trivial explanation in genomes \citep{Ferrer2009e,Wilde1989a}. Claiming that the scaling defined by Eq.\ \ref{Soles_law_equation} is inevitable \citep{Sole2010a} is equivalent to claiming that $Y$ must be mean independent of $X$ in any circumstance, a very strong requirement for real language, music and genomes.
Eq.\ \ref{Soles_law_equation} should be regarded as a baseline instead of a fact for the relationship between chromosome number and mean chromosome length. Rather than signs of complexity, certain power laws (the value of the exponent can be crucial) might indicate the control that must be considered before any claim of "sufficient biological complexity" can be made.

\section*{Acknowledgements}

This article has benefited enormously from the comments of anonymous reviewers.
We are grateful to P. Delicado, R. Gavald\`a and E. Pons for their valuable mathematical insights. We owe the counterexample showing that uncorrelation does not imply mean independence to P. Delicado. We are also grateful to G. Bel-Enguix and N. Forns for helpful discussions. This work was supported by the grant {\em Iniciaci\'o i reincorporaci\'o a la recerca} from the Universitat Polit\`ecnica de Catalunya, the grants BASMATI (TIN2011-27479-C04-03) and OpenMT-2 (TIN2009-14675-C03) from the Spanish Ministry of Science and Innovation and the grant 2/0038/12 from the VEGA funding agency (JM).

\bibliographystyle{DeGruyter}
\bibliography{../../../biblio/complex,../../../biblio/rferrericancho,../../../biblio/ling,../../../biblio/cs,../../../biblio/maths}

\end{document}